\documentclass[a4paper,aps,pre,10pt,floatfix,twocolumn,showpacs,superscriptaddress,nofootinbib]{revtex4-1}

\usepackage{subfigure}
\usepackage{amssymb}
\usepackage{amsfonts}
\usepackage{amsmath}
\usepackage{amsthm}
\usepackage{epsfig}
\usepackage{graphicx}
\usepackage[normalem]{ulem}
\usepackage[usenames,dvipsnames]{color}
\usepackage[latin1]{inputenc}
\usepackage[pdftex]{hyperref}
\usepackage{comment}
\usepackage{ifpdf}
\usepackage{setspace}

 \newcommand{\revis}[1]{{\color{black} {#1}}}
\newcommand{\lrangle}[1]{\langle{#1}\rangle}

\begin{document}
\title{Multiple transitions of
the susceptible-infected-susceptible epidemic model
on complex networks}
\author{Ang\'elica S. Mata}
\author{Silvio C. Ferreira}
%\email{silviojr@ufv.br}
\affiliation{Departamento de F\'isica, Universidade Federal de Vi\c cosa,
36570-000, Vi\c cosa, MG, Brazil}
\begin{abstract}
The epidemic threshold of the susceptible-infected-susceptible (SIS) dynamics on 
random networks having a power law degree distribution with exponent $\gamma>3$ 
has been investigated using different mean-field approaches, which predict 
different outcomes. We performed \revis{extensive simulations} in the 
quasistationary state for a comparison with these mean-field theories. We 
observed concomitant multiple transitions in \revis{individual networks} presenting large 
gaps in the degree distribution and the obtained multiple epidemic thresholds 
are well described by different mean-field theories. We observed that the 
transitions involving thresholds which vanishes at the thermodynamic limit 
involve localized states, in which a vanishing fraction of the network 
effectively contribute to epidemic activity,  whereas an endemic state, with a 
finite density of infected vertices, occurs at a finite threshold. The multiple 
transitions are related to the activations of distinct sub-domains of the 
network, which are not {directly} connected. %revised
\end{abstract}


\maketitle

\section{Introduction}

Phase transitions involving equilibrium and non-equilibrium processes on  
complex networks have begun drawing an increasing interest soon after 
the boom of network science at late 90's~\cite{Dorogovtsev08}. 
Percolation~\cite{albert2000}, epidemic spreading~\cite{Pastor01,Pastor01b}, 
and spin systems~\cite{DoroIsing} are only a few examples of breakthrough in 
the investigation of critical phenomena in complex networks. Absorbing 
state phase transitions~\cite{henkel08}  have become a paradigmatic issue in 
the interplay between nonequilibrium systems and  complex 
networks~\cite{Castellano08,Hong2007,JuhaszCP,Mata14,cpannealed}, being the 
epidemic spreading a prominent example where high complexity emerges from very 
simple dynamical rules on heterogeneous 
substrates~\cite{Pastor01,Pastor01b,Castellano10, Goltsev12, Ferreira12, 
boguna2013nature, Lee2013}.

The existence or absence of finite epidemic thresholds involving an endemic phase
of the susceptible-infected-susceptible (SIS) model on scale-free
(SF) networks with a degree distribution $P(k)\sim k^{-\gamma}$, where $\gamma$
is the degree exponent, has been target of a recent and intense
investigation~\cite{Castellano10,Cator12a,vanMieghem09, Goltsev12, Ferreira12,
boguna2013nature, Lee2013}. In the SIS epidemic model,
individuals can only be in one of two states: infected or susceptible. Infected
individuals become spontaneously healthy at rate $1$ (this choice fixes the time
scale), while the susceptible ones are infected at rate $\lambda n_i$, where
$n_i$ is the number of infected contacts of a vertex $i$.

Distinct theoretical approaches for the SIS model were devised  to determine an 
epidemic threshold $\lambda_c$ separating an absorbing, disease-free state from 
an active phase~\cite{Castellano10,Cator12a,vanMieghem09, Goltsev12, Ferreira12, 
boguna2013nature, Lee2013, Chakrabarti08,mata2013pair}. The quenched mean-field 
(QMF) theory~\cite{Chakrabarti08} explicitly includes the entire structure of 
the network through its adjacency matrix while the heterogeneous mean-field 
(HMF) theory~\cite{Pastor01,Pastor01b} performs a coarse-graining of the network 
grouping vertices accordingly their degrees. The HMF theory predicts a vanishing 
threshold for the SIS model for the range $2< \gamma\le 3$ while a finite 
threshold is expected for $\gamma>3$. Conversely, the QMF theory states a 
threshold inversely proportional to the largest eigenvalue of the adjacency 
matrix, implying that the threshold vanishes for any value of 
$\gamma$~\cite{Castellano10}. However, Goltsev \textit{et al}.~\cite{Goltsev12} proposed 
that QMF theory predicts the threshold for an endemic phase, in which a finite 
fraction of the network is infected, only if the principal eigenvector of 
adjacency matrix is delocalized. In the case of a localized principal 
eigenvector, that usually happens for large random networks with 
$\gamma>3$~\cite{odor2014localization}, the epidemic threshold is associated to 
the eigenvalue of the first delocalized eigenvector. For $\gamma<3$, there 
exists a consensus for SIS thresholds: both HMF and QMF are equivalent and 
accurate for $\gamma<2.5$ while QMF works better for 
$2.5<\gamma<3$~\cite{Ferreira12,mata2013pair}.

Lee {\it et al.}~\cite{Lee2013} proposed that for a range $\lambda_{c}^{QMF} < 
\lambda < \lambda_c$ with a nonzero $\lambda_c$, the hubs in a random network 
become infected generating epidemic activity in their neighborhoods but 
high-degree vertices produce independent active domains only when they are not 
directly connected. {These independent domains were classified as rare-regions, 
in which activity can last for very long times (increasing exponentially with 
the domain size~\cite{Noest}), generating Griffiths phases 
(GPs)~\cite{Noest,Vojta}.} The sizes of these active domains increase for 
increasing $\lambda$  leading to the overlap among them and, finally, to an 
endemic phase for $\lambda > \lambda_c$. However, on networks where almost all 
hubs are directly connected, it is possible to sustain an endemic state even in 
the limit $\lambda \rightarrow 0$ due to the mutual reinfection of connected 
hubs. Inspired in the appealing arguments of Lee {\it et al.}~\cite{Lee2013}, 
Bogu\~n\'a, Castellano and Pastor-Satorras (BCPS)~\cite{boguna2013nature} 
proposed a semi-analytical approach taking into account a long-range reinfection 
mechanism and found a vanishing epidemic threshold for $\gamma>3$. They 
compared their theoretical predictions with simulations starting from a single 
infected vertex and a diverging epidemic lifespan was used as a criterion to 
determine the thresholds. However, the applicability of BCPS theory to determine 
a phase transition involving an endemic phase has been 
debated~\cite{Lee2013comment,*boguna2014reply}.

In this work, we performed extensive simulations and found  that the SIS 
dynamics on SF networks  with exponent $\gamma>3$ can exhibit multiple 
transitions, with multiple thresholds, \revis{which are clearly resolved when 
the degree distribution presents outliers separated by large gaps. These gaps 
permits the formation of non directly connected domains centered on hubs with 
different connectivity  and thus having distinct local activation thresholds}. 
Thresholds consistent with those predicted by QMF, HMF and BCPS theories were 
found in our analysis. Moreover, our finds indicate that the vanishing 
thresholds, as those predicted by QMF~\cite{Castellano10} and BCPS 
theories~\cite{boguna2013nature}, involve long-term  but still localized 
epidemics rather than an endemic state, in which a finite fraction of the 
network has non-vanishing probability to be infected in the thermodynamic limit. 
We propose that this localized long-term epidemics takes place in  domains 
involving a \revis{few} hubs with very large degree and their nearest-neighbors. 
Finally, our numerical results show a transition to the endemic state occurring 
at a finite threshold, which is intriguingly well described  by the classic and 
simpler HMF theory~\cite{Pastor01,Pastor01b}. 

Our paper is organized as follows: in Sec. \ref{sec:simulations} we present 
simulation procedures, discuss important technical details of the  quasistationary (QS) 
method used in this work and provide a comparison between QS method and the  
lifespan simulation method proposed in Ref.~\cite{boguna2013nature}. Section 
\ref{sec:results} is devoted to describe  the numerical results obtained from QS 
simulations and in  Sec.~\ref{sec:conclusions} we draw our concluding 
remarks. Finally,  an example where the lifespan method does not determine the 
endemic phase in systems with multiple transitions while the QS method does is 
presented in Appendix \ref{sec:appendix}. 

\section{Simulation methods}
\label{sec:simulations}

We implement the SIS model using a \revis{modified Gillespie simulation 
scheme~\cite{Gillespie} provided in Ref.~\cite{Ferreira12}}:  At each time step, 
the number of infected nodes $N_i$ and edges emanating from them $N_k$ are 
computed and time is incremented by\footnote{In the original Gillespie algorithm 
for the simulation of stochastic processes~\cite{Gillespie}, the time increment 
is drawn from an exponential distribution with mean $dt$. However, this 
stochasticity in time increment did not play an important rule in our analysis 
due to the large averaging used.} $\Delta t = 1/(N_i+\lambda N_k)$.  With probability 
$N_i/(N_i+\lambda N_k)$ one infected node is selected at random and becomes 
susceptible. With the complementary probability $\lambda N_k/(N_i+\lambda N_k)$ 
an infection attempt is performed in two steps: (i) A infected vertex $j$ is 
selected with probability proportional to its degree. (ii) A nearest neighbor of 
$j$ is selected with equal chance and, if susceptible, is infected. If the 
chosen neighbor is infected nothing happens and simulation runs to the next time 
step. Notice that $\lambda N_k$ is the total infection rate emanating from 
infected vertices and  the frustrated attempts compensate this exceeding rate. 
\revis{The frustrated attempts constitute the central alteration in relation to original 
Gillespie algorithm.} The  numbers of infected nodes and related links are 
updated accordingly, and the  whole process is iterated. 

The simulations were performed using the QS method 
~\cite{DeOliveira05,cpannealed} that, to our knowledge, is the most robust 
approach to overcome the difficulties intrinsic to the stationary simulations of 
finite systems with absorbing states. In this method, every time the system 
tries to visit an absorbing state it jumps to an active configuration previously 
visited during the simulation (a new initial condition). {This method  
reproduces very accurately the standard QS method where averages are performed 
only over samples that did not visit the absorbing 
state~\cite{DeOliveira05,dickman2002quasi} and its convergence to the real QS 
state was proved~\cite{blanchet}.} To implement the method, a list containing 
$M=70$ configurations is stored and constantly updated. The updating is done by 
randomly picking up a stored configuration and replacing it by the current one 
with probability $p_r\Delta t$.  We fixed $p_r\simeq 10^{-2}$ since no 
significant dependence on this parameter was detected for a wide range of 
simulation parameters. After a relaxation time $t_r$, the averages are computed 
over a time $t_{av}$.

The characteristic relaxation time is always short for epidemics on random 
networks due to the very small average shortest path~\cite{Newman10}. Typically, 
a QS state is reached for $t>10^4$ for the simulation parameters investigated. 
So, we used $t_r=10^5$. The averaging time, on the other hand, must be large 
enough to guaranty that epidemics over the whole network was suitably averaged. 
It means that very long times are required for very low QS density (sub-critical 
phase in phase transition jargon) whereas relatively short times are sufficient 
for high density states. Since long times are computationally prohibitive for 
highly infected QS states, we used averaging times from $10^5$  to $10^9$, 
being the larger the average time the smaller the infection rate. Notice that 
the simulation time step becomes tiny for a very supercritical system (large 
number of infected vertices) and a huge number of configurations are visited 
during a unity of time. It is important to notice that the QS method becomes 
expendable for a large part of our simulations since the system never visits the 
absorbing state for the considered simulation times.

Both equilibrium and non-equilibrium critical phenomena are hallmarked by 
simultaneous diverging correlation length and time, which microscopically 
reflect the divergence of the spatial and temporal fluctuations~\cite{henkel08}, 
respectively. Even tough a diverging correlation length has little sense on 
complex networks due to the small-world property~\cite{watts98}, the diverging 
temporal fluctuation concept is still applicable. We used different criteria to 
determine the thresholds, relied on the fluctuations or singularities of the 
order parameters, as explained below.

The QS probability $\bar{P}(n)$, defined as the probability that the system has 
$n$ occupied vertices in the QS regime, is computed during the averaging time 
and basic QS quantities, as lifespan and density of infected vertices, are 
derived form $\bar{P}(n)$~\cite{DeOliveira05}. Thus, thresholds for finite 
networks can be estimated using the modified susceptibility~\cite{Ferreira12}
\begin{equation}
 \chi\equiv\frac{\lrangle{n^2}-\lrangle{n}^2}{\lrangle{n}}=
\frac{N(\lrangle{\rho^2}-\lrangle{\rho}^2)}{\lrangle{\rho}},
\label{eq:chi}
\end{equation}
that does exhibit a  divergence at the transition point for 
SIS~\cite{Ferreira12,mata2013pair,Lee2013}  and contact 
process~\cite{RonanEPJB,Mata14} models on networks. The choice of the 
alternative definition, Eq.~(\ref{eq:chi}), instead of the standard 
susceptibility 
$\tilde{\chi}=N(\lrangle{\rho^2}-\lrangle{\rho^2})$~\cite{henkel08} is due to 
the peculiarities of dynamical processes on the top of  complex 
networks\footnote{See discussion in Ref.~\cite{Mata14}, section 3.}.

\begin{figure}[hbt!]
 \centering
 \includegraphics[width=6.9cm]{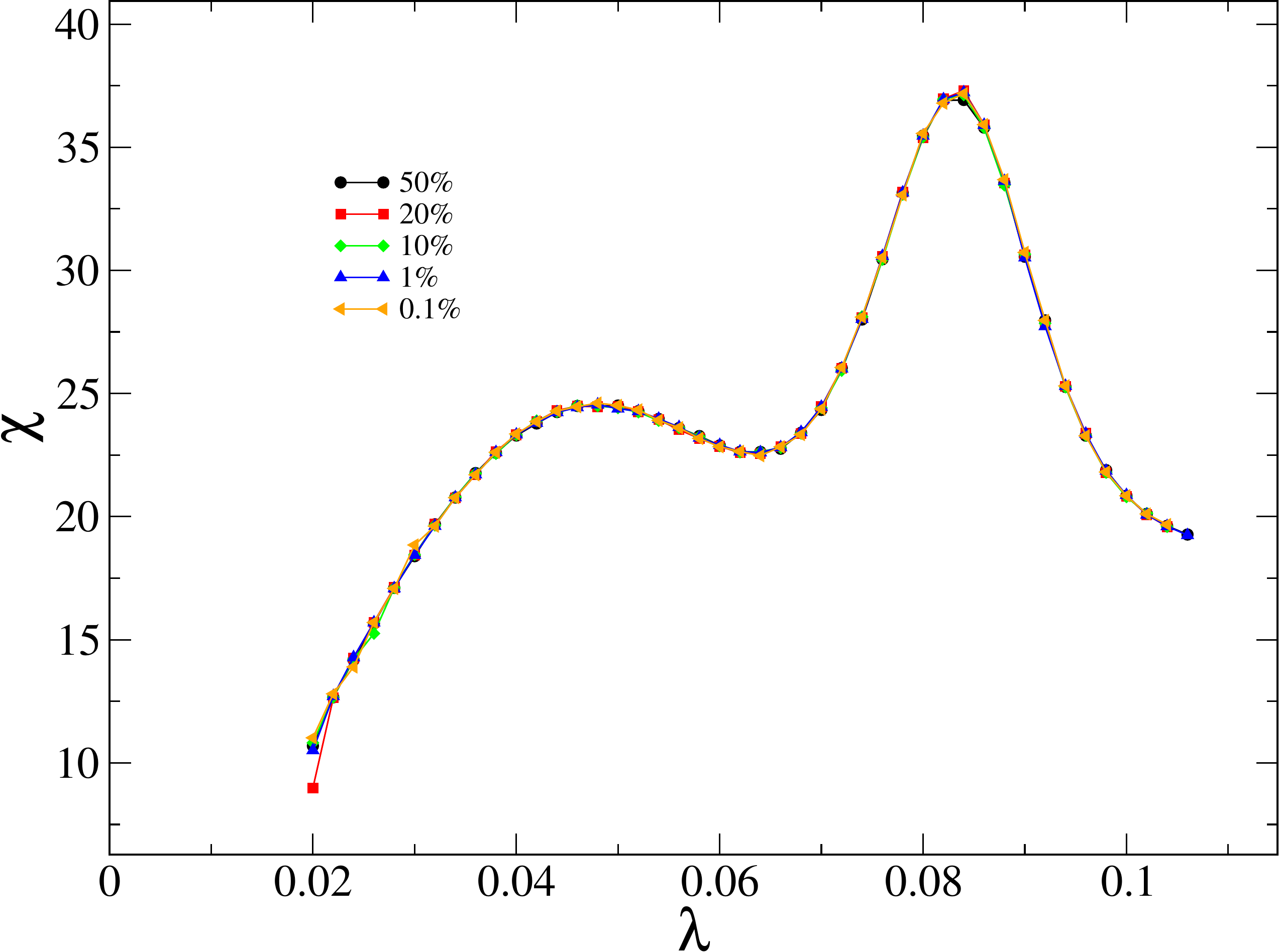}
 % g350N7_IC.pdf: 733x546 pixel, 72dpi, 25.86x19.26 cm, bb=0 0 733 546
 \caption{(Color online) Susceptibility against infection rate for SIS model on {a 
single network} with different fraction of initially infected vertices, 
{which are randomly distributed in the network}. The network 
parameters are $\gamma=3.5$, $k_0=3$ and $N=10^6$.}
 \label{fig:IC}
\end{figure}

It is expected that the QS state does not depend on the initial condition. 
Figure~\ref{fig:IC} shows a comparison of QS simulations for the same network 
with different initial densities $\rho(0)=10^{-3}$ to 0.5, randomly distributed. The network was 
generated with the uncorrelated configuration model (UCM)~\cite{Catanzaro05}, 
where vertex degree is selected from a power-law distribution\footnote{To 
generate the degree distribution we used the improved rejection method provided 
in Ref.~\cite{NRE}.} with a lower bound $k_0=3$. The results are  
independent of the initial condition. Also, the QS method was compared with the 
so-called  $\epsilon$-SIS model~\cite{Cator12} where  a small rate $\epsilon$ of 
spontaneous infection is assumed for each network vertex. The thresholds 
involving long-term epidemics are the same as those of the QS method~\cite{sander}.

Reference~\cite{boguna2013nature} claimed that the QS method is 
unreliable\footnote{In private communications, authors of 
Ref.~\cite{boguna2013nature} clarified that the multiple peaks observed in the 
susceptibility curves cannot unambiguously define the lifespan divergence. 
However, they passed over the fact that a lifespan is easily extracted from QS 
simulations using Eq.~(\ref{eq:tauqs}).} for networks with degree exponents 
$\gamma>3$ and proposed a new simulation strategy, which is referred here as 
lifespan simulation method. In order to draw a comparison with the QS method, we 
implemented the lifespan method exactly as in Ref.~\cite{boguna2013nature}: The 
simulation starts with a single infected vertex located at the most connected 
vertex of the network and stops when  either the system visits the absorbing 
state or 50\% of all vertices (the epidemic coverage) were infected at least 
once along the simulation. The duration of the epidemic outbreak is computed and 
only runs that visited the absorbing state are used to compute the average 
lifespan since those that reached 50\% of coverage are assumed as having an 
infinite lifespan.  {The number of runs varies from $10^3$, for  largest $N$ and 
$\lambda$, to $10^6$, for the smallest $\lambda$.}

We applied both methods to the SIS model on UCM networks with $\gamma=3.50$, 
minimum degree~$k_0=3$, and upper cutoff $k_{max}=\lrangle{k_{max}}$, {in which 
$\lrangle{k_{max}}$  is the analytically determined mean value  of the most 
connected vertex $k_{max}$ of a random degree sequence with distribution $P(k)$ 
without upper bounds}, to compare with the results of 
Ref.~\cite{boguna2013nature}. The constraint $k_{max}=\lrangle{k_{max}}$ avoids 
fluctuations in the most connected vertex and, consequently, in the largest 
eigenvalue of the adjacency matrix and is useful for comparisons with the QMF 
theory~\cite{Ferreira12}. We remark that the constraint 
$k_{max}=\lrangle{k_{max}}$ is only used in this comparison.
\begin{figure}[htb!]
 \centering
 \includegraphics[width=8.5cm]{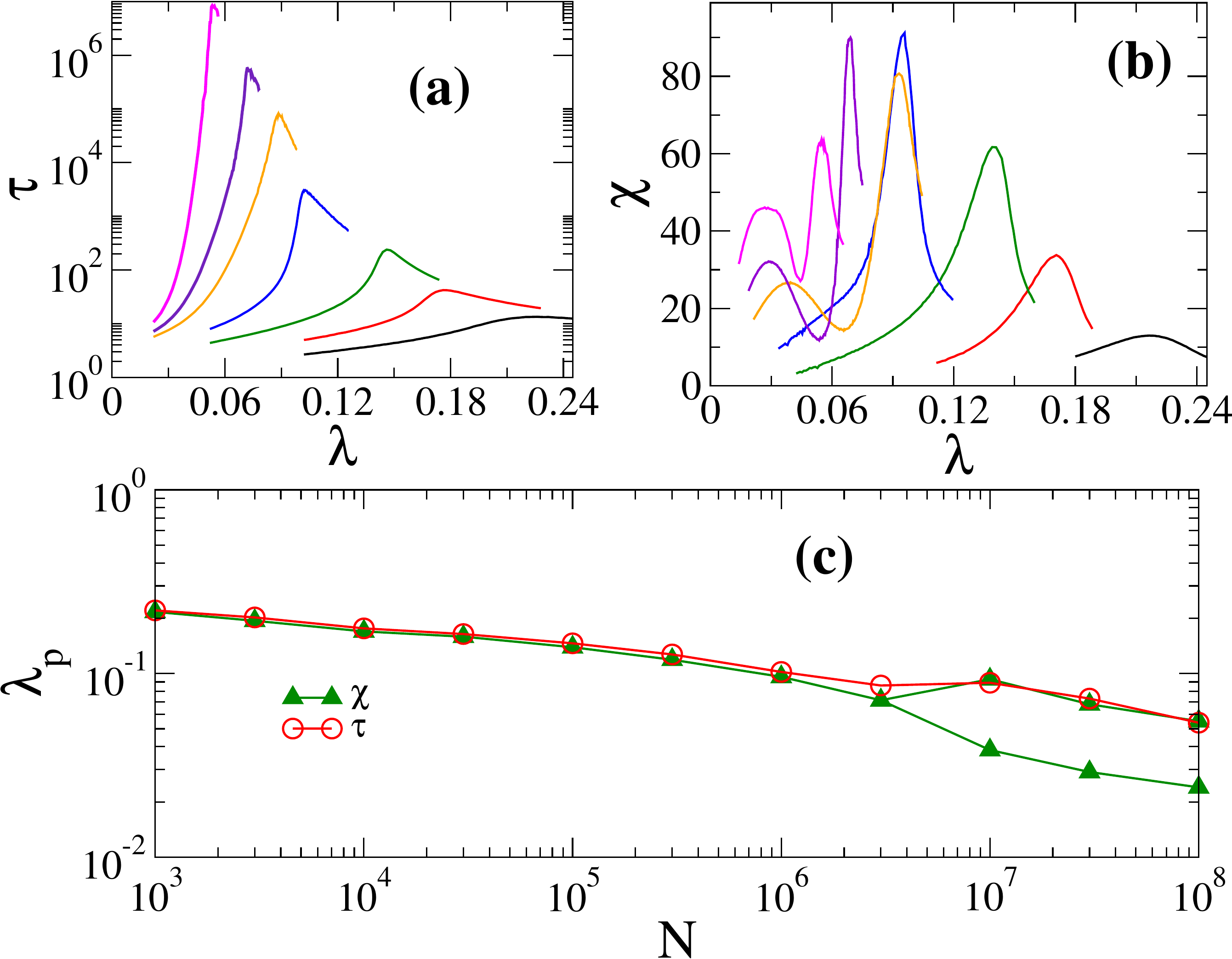}
 % taug350.pdf: 746x556 pixel, 72dpi, 26.32x19.61 cm, bb=0 0 746 556
 \caption{(Color online) Numerical determination of the thresholds for the SIS 
model on  UCM networks with $\gamma=3.50$, $k_0=3$ and 
$k_{max}=\lrangle{k_{max}}$ for network sizes 
$N=10^3,10^4,10^5,10^6,10^7,3\times 10^7$, and $10^8$, increasing from the 
right. {A same network sample for each size was used in both methods.} Both (a) 
lifespan calculated using the method of Ref.~\cite{boguna2013nature} and (b) 
susceptibility via QS method are shown in the top panels. (c) Peak positions as 
functions of the network size estimated with both 
methods. }
 \label{fig:sustau_sis}
\end{figure}

Figures~\ref{fig:sustau_sis}(a) and (b) show the lifespan and susceptibility 
against infection rate for networks of different sizes. The peak positions 
against network size are compared in Fig~\ref{fig:sustau_sis}(c). As can be 
clearly seen, the right susceptibility peaks are very close to the lifespan ones 
showing that the susceptibility method is able to capture the same transitions 
as the lifespan method does but going beyond as discussed in the rest of the 
paper. It is worth noticing that if larger values of $\lambda$ are simulated, 
other peaks will emerge in susceptibility curves even using the cutoff $ k \le 
\langle k_{max}\rangle$. These multiple peaks were not reported in previous 
works dealing with the same network 
model~\cite{Ferreira12,mata2013pair,boguna2013nature}.

Moreover, a lifespan is also obtained in the QS 
method as~\cite{DeOliveira05} 
\begin{equation}
 \tau_{qs}=\frac{1}{\bar{P}(1)}.
 \label{eq:tauqs}
\end{equation}
We checked that the lifespans obtained in the QS method and those of 
Ref.~\cite{boguna2013nature} diverge around the same threshold; the basic  
difference is that the former is ``infinite'' above the threshold whereas the 
latter remains finite.  

In a partial summary, we verified that the lifespan method predicts an epidemic 
threshold when an activity survives for long times, but there  is no guaranty 
that it  is necessarily an endemic phase (see appendix~\ref{sec:appendix} for a 
concrete counter-example). On the other hand, the QS analysis is able to 
simultaneously determine transitions involving endemic as well as localized 
states and the one involving a diverging lifespan is resolved using 
Eq.~(\ref{eq:tauqs}). So, we conclude that lifespan method must not be used 
alone in systems with multiple transitions since it captures the first 
transition with a long-term activity.

\section{Numerical Results}
\label{sec:results}
Two-peaks on susceptibility against infection rate for SIS  were firstly 
reported in Ref.~\cite{Ferreira12}, which  focused on the analysis of the peak 
at low $\lambda$ and showed that it is well described by the QMF theory (see 
also Ref.~\cite{mata2013pair}) but did not realize that the peak at higher 
$\lambda$ is the one associated to a diverging lifespan 
(Fig.~\ref{fig:sustau_sis}). However, depending on the network realization, the 
susceptibility curves can exhibit much more complex behaviors with multiple 
peaks for values of $\lambda$ larger than those reported in 
Refs.~\cite{Ferreira12,mata2013pair}. These complex behaviors become very 
frequent for large networks. From now on, we scrutinize such a complex behavior 
to unveil its origin and implications to the epidemic activity.
\begin{figure}[ht]
 \centering
 \includegraphics[width=8.5cm]{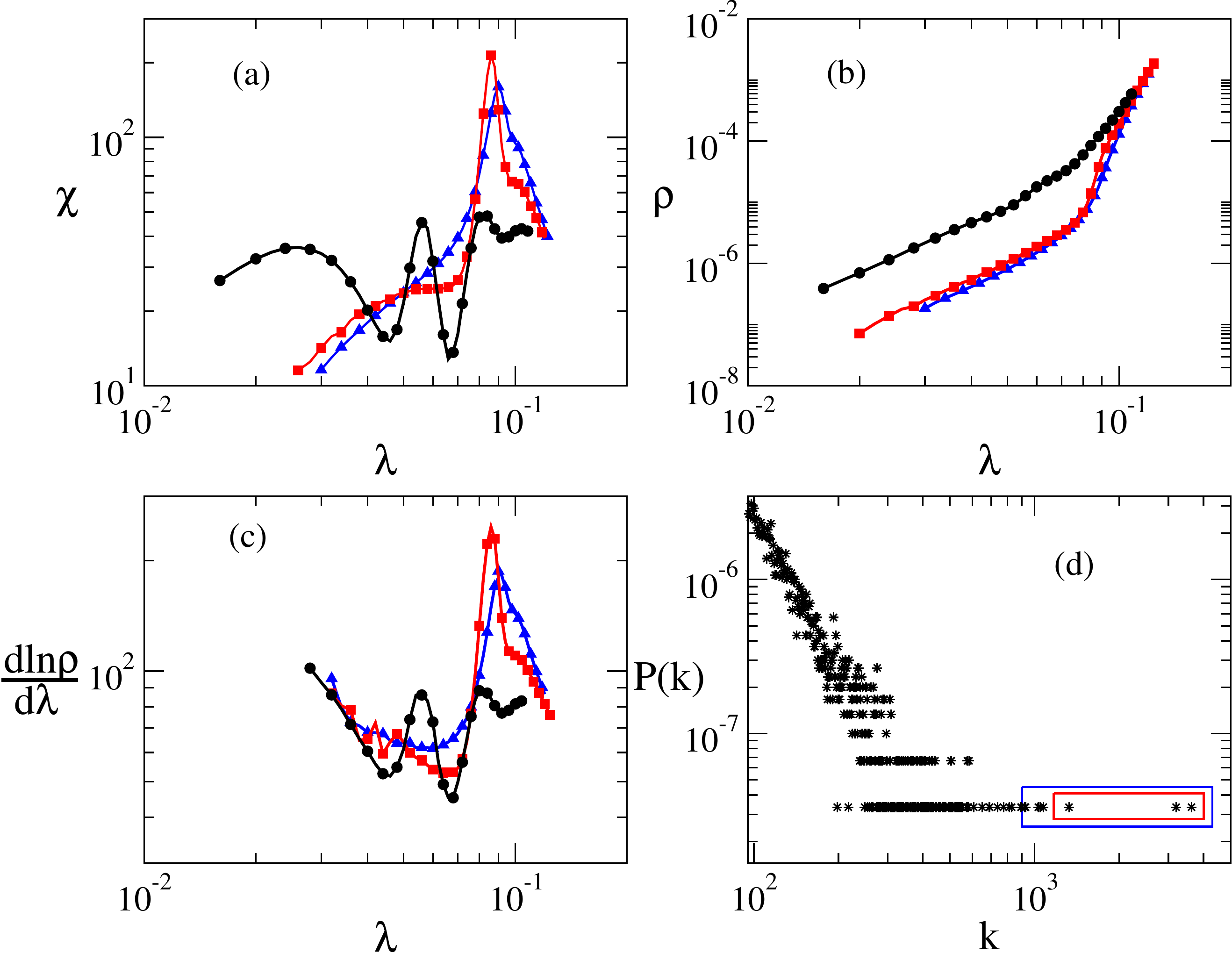}
 % chig350N10a1.pdf: 730x557 pixel, 72dpi, 25.75x19.65 cm, bb=0 0 730 557
 \caption{(Color online) (a) Susceptibility, (b) stationary density and (c) its logarithmic
derivative versus infection rate for a SF network with $3\times 10^7$ vertices,
degree exponent $\gamma=3.5$, minimum degree $k_0=3$ and $k_{max}$ unconstrained. The
degree distribution is shown in panel (d). Different immunization strategies are
shown: Black circles represent no immunization; red squares represent the
immunization of three largest outliers (inner box in panel (d)); blue triangles
represent the immunization of eight most connected vertices (outer box in panel
(d)). \label{fig:chiam1}}
\end{figure}

Figure~\ref{fig:chiam1}(a) shows a typical susceptibility curve (black) 
exhibiting such a complex behavior for an UCM network. The degree distribution 
is shown in Fig.~\ref{fig:chiam1}(d). 
{Multiple peaks are observed {only if} the 
degree distribution exhibits a few large gaps, in particular in the tails.
These few  vertices\footnote{The number of 
outliers can estimated as $N\int_{k \gtrsim \lrangle{k_{max}}}P(k) dk\sim \mathcal{O}(1)$.} 
with degree $k \gtrsim \lrangle{k_{max}}$
%, where $k_{max}$ is the maximum 
%value obtained in the generation  of $N$ random variables with distribution 
%$P(k)$, 
are hereafter called outliers.}
%{For example, for $\gamma=3.1$ and $N=10^8$ the distribution 
%possesses many outliers but neither large gaps or multiple peaks were observed 
%in all samples we checked. Notice these gaps and multiple peaks will possibly 
%emerge if one could simulate much  larger sizes (see discussion below).}
Notice that the multiple peaks are 
not detected by the lifespan simulation method~\cite{boguna2013nature}. 
The role played by outliers is evidenced by their immunizations\footnote{Immunized 
vertices cannot be infected, which is equivalent to removing them from the 
network.} as illustrated in Fig.~\ref{fig:chiam1}. For instance, the 
immunization of the three most connected vertices is sufficient to destroy two peaks 
and to enhance others. The stationary density varies abruptly close to the 
thresholds determined via susceptibility peaks, Figs.~\ref{fig:chiam1}(b) and 
(c), which  is an  evidence of the singular behavior of the order parameter 
$\rho$.

{The presence of gaps is a characteristics of large degree 
sequences with a power law distribution. The statistical representativity of 
specific properties of a finite set of networks,  generated under the same conditions, 
in relation to the entire ensemble is a complex issue~\cite{delGenio}, but the 
existence of gaps can be understood with a simple non-rigorous reasoning. Using 
extreme value theory one can show that the most connected vertex has an average  
$\lrangle{k_{max}}\sim N^{1/(\gamma-1)}$~\cite{Boguna09}. However, this mean 
value is not representative of the highest degree since the dispersion 
$\sigma_{max}=\sqrt{\lrangle{k_{max}^2} -\lrangle{k_{max}}^2}$  diverges 
as\footnote{This result can be 
derived using the same steps to obtain $\lrangle{k_{max}}$ 
in~Ref.~\cite{Boguna09}.}
$\sigma_{max}\sim N^{1/(\gamma-1)}$ for $\gamma>3$.
Outliers should behave in this same way and therefore we 
expect larger dispersion in outlier connectivity as  network size increases.
}

It is interesting to observe that while the peaks at small $\lambda$ 
can or not appear depending on the presence of outliers and  gaps, the 
rightmost one essentially does not change its position from a network 
realization to another, {such that it should depend on network properties
representative of the entire ensemble of networks with a specified 
set of parameters. Indeed, later we will see that the behavior of the rightmost peak 
is qualitatively described by the HMF threshold which depends only on $\lrangle{k^2}$
and $\lrangle{k}$.}

A deeper physical explanation for the multiple peaks can be extracted using
another order parameter in the QS state, the participation ratio (PR), defined as
\begin{equation}
 \Phi=\frac{1}{N}\frac{(\sum_i\rho_i)^2}{\sum_i\rho_i^2},
\end{equation}
where $\rho_i$ is the probability that the vertex $i$ is infected in the 
stationary state. The inverse of the PR is a standard measure for 
localization/delocalization of  states in condensed matter~\cite{Bell} and has 
been applied to statistical physics problems~\cite{Plerou} including epidemic 
spreading on networks~\cite{Goltsev12,Bathelemy,odor2013spectral}. The limiting 
cases of totally delocalized ($\rho_i=\rho$~$\forall~i$) and localized 
($\rho_i=\rho\delta_{i,0}$ where 0 is  the vertex where localization occurs) 
states are $\Phi = 1$ and $\Phi = 1/N$, respectively.

\begin{figure}[hbt]
 \centering
 \includegraphics[width=7.0cm]{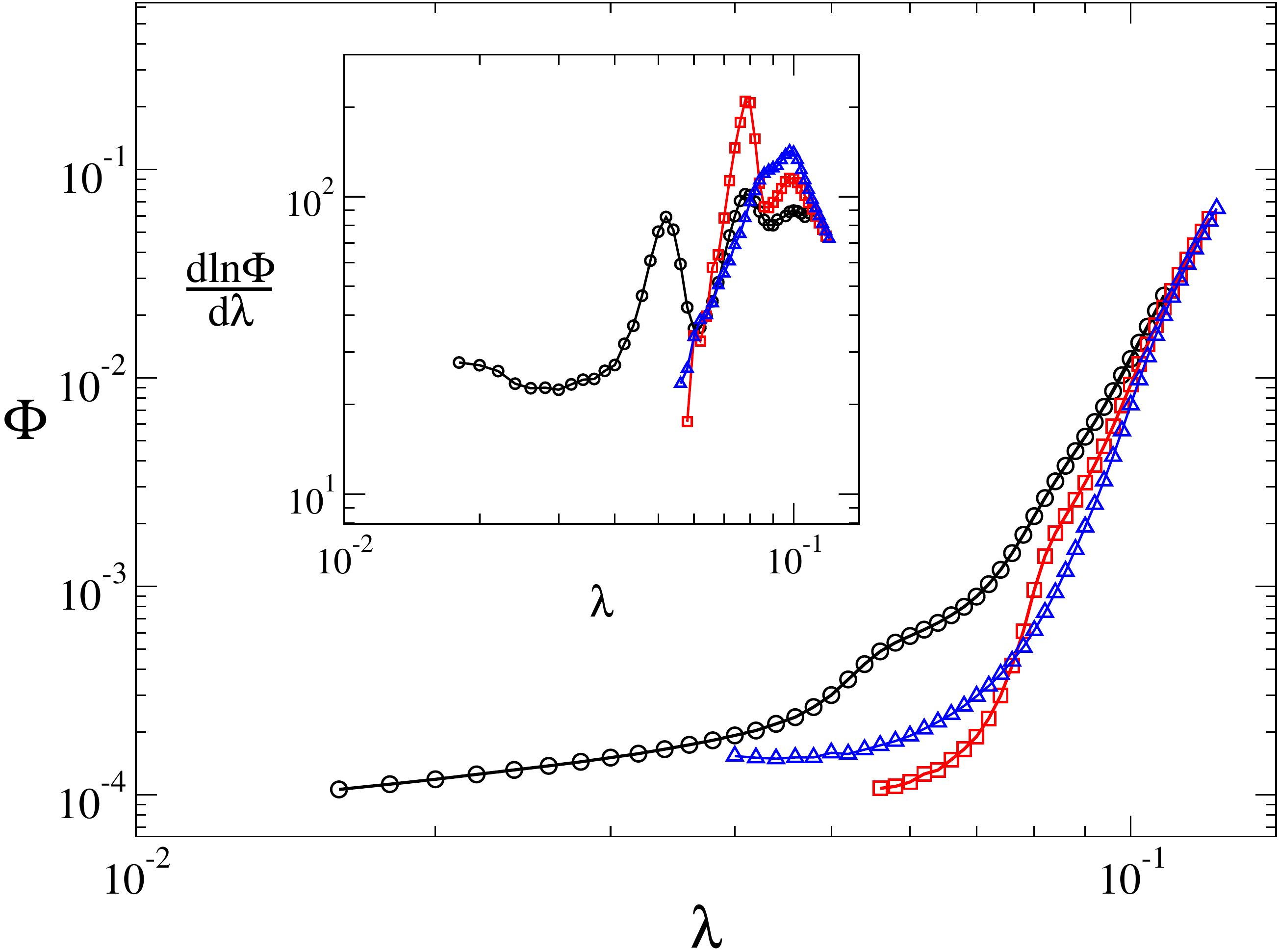}
\caption{(Color online) Main: PR as a function of the infection rate for the same network and 
immunization strategies as in Fig.~\ref{fig:chiam1}. Symbols as in 
Fig.~\ref{fig:chiam1}. Inset: Logarithmic derivative of the PR as 
a function of the infection rate. \label{fig:Phi}}
\end{figure}

The PR as a function of the infection rate is shown in  Fig.~\ref{fig:Phi}. The 
PR is an estimate of the fraction of vertices that effectively contribute to the 
present epidemic activity. Thus, the multiple transitions are related to the 
rapid delocalization processes of the epidemics as $\lambda$ increases, 
hallmarked by the singular behavior of $\Phi$ around distinct values of 
$\lambda$. When the PR corresponds to a finite fraction of the network in an 
active phase one has an authentic endemic state, since a finite fraction of 
nodes has a non-vanishing probability of being infected at the same time. 
The logarithmic derivative of the PR exhibits several peaks in analogy to
susceptibility peaks, as shown in the inset of Fig.~\ref{fig:Phi}. Indeed, PR can be
seen as a susceptibility but  from an origin different of $\chi$. The latter
is a measure of stochastic fluctuations of the order parameter (density of
infected vertices) whereas the former is measure of stationary spatial
fluctuations that make sense only for heterogeneous substrates.

\begin{figure}[hbt]
 \centering
 \includegraphics[width=8.5cm]{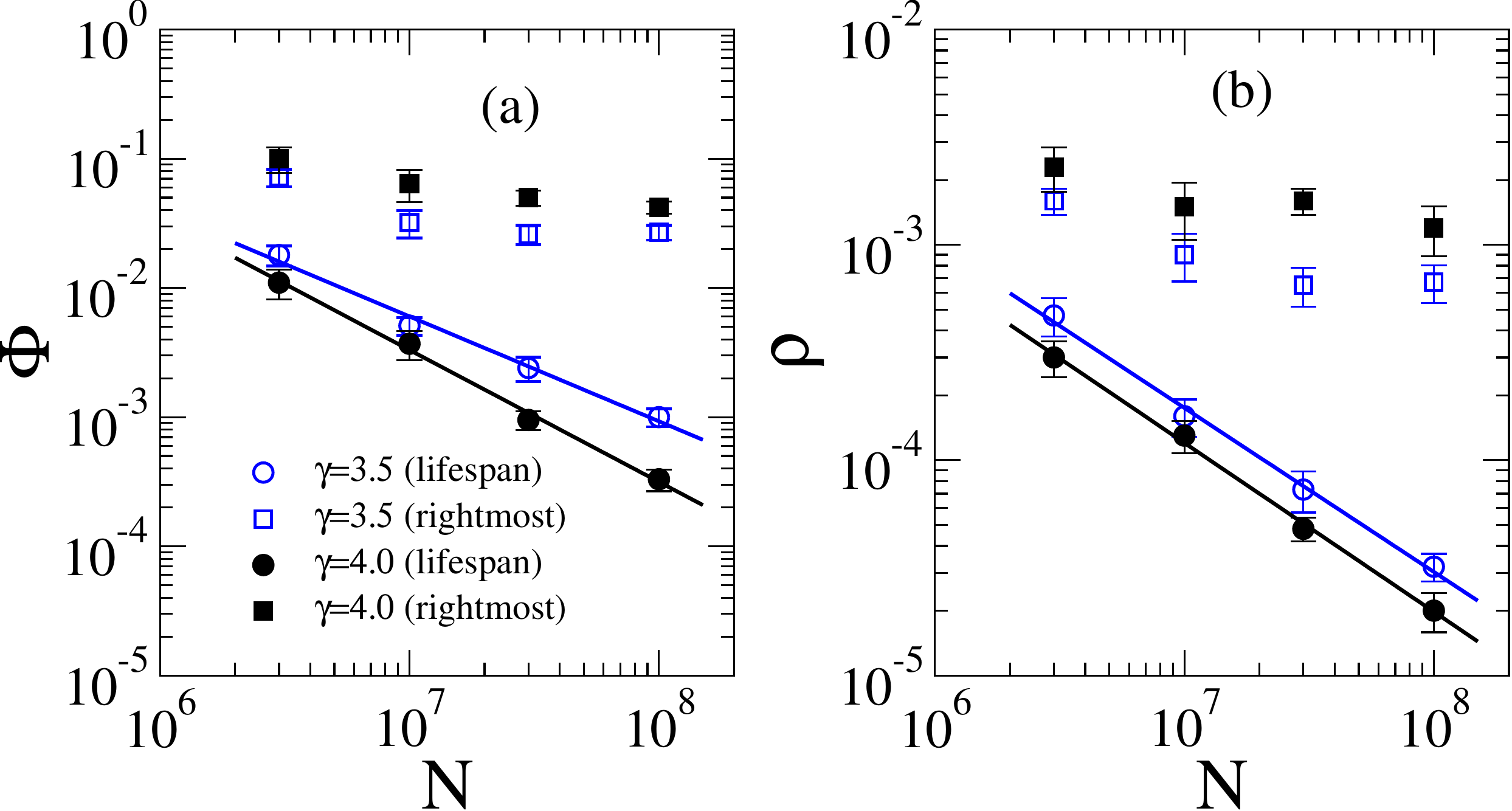}
 %\subfigure[][]{\includegraphics[width=4.1cm]{./rho2.pdf}}
 \caption{(Color online) (a) PR against size for a fixed distance $\lambda-\lambda_p=0.012$ to 
either lifespan (circles) and rightmost (squares) peaks. (b) The same analysis of 
panel (a) for QS density. Lines are power regressions. {At least 10  network 
samples were used to perform averages for $\lambda>\lambda_p^{right}$ (top curves)
and at least 20 for $\lambda_p^{ls}<\lambda<\lambda_p^{right}$ (bottom curves).}}
 \label{fig:PRsup}
\end{figure}

The PR against network size for a fixed distance to either $\lambda_p^{ls}$ (the 
threshold marking the lifespan divergence) and $\lambda_p^{right}$  (the 
threshold referent to the rightmost peak observed for susceptibility) are shown 
in Fig.~\ref{fig:PRsup}(a). In the presented size range, the PR decays as a power 
law for a fixed distance to the lifespan peaks. Analogous results are obtained for $\bar{\rho}$ vs $N$ curves (see Fig. 
\ref{fig:PRsup}(b)). The power 
regressions yield  approximately $\Phi\sim N^{-0.8}$ and $N^{-1}$ for 
$\gamma=3.5$ and 4, respectively, $\rho\sim N^{-0.8}$ for both $\gamma=3.5$ 
and 4. These decays constitute a strong evidence for epidemics localization at 
$\lambda\gtrsim \lambda_p^{ls}$ whereas the constant dependence on $N$ observed for 
$\lambda\gtrsim \lambda_p^{right}$  represents an endemic 
phase\footnote{Notice that a scaling $\bar{\rho}\sim (\lambda-\lambda_p)^\beta$, 
independent of the size, is expected for an usual endemic phase transition in the 
thermodynamic limit~\cite{henkel08}.}. 

\begin{figure}[hbt]
 \centering
 \includegraphics[width=8.0cm]{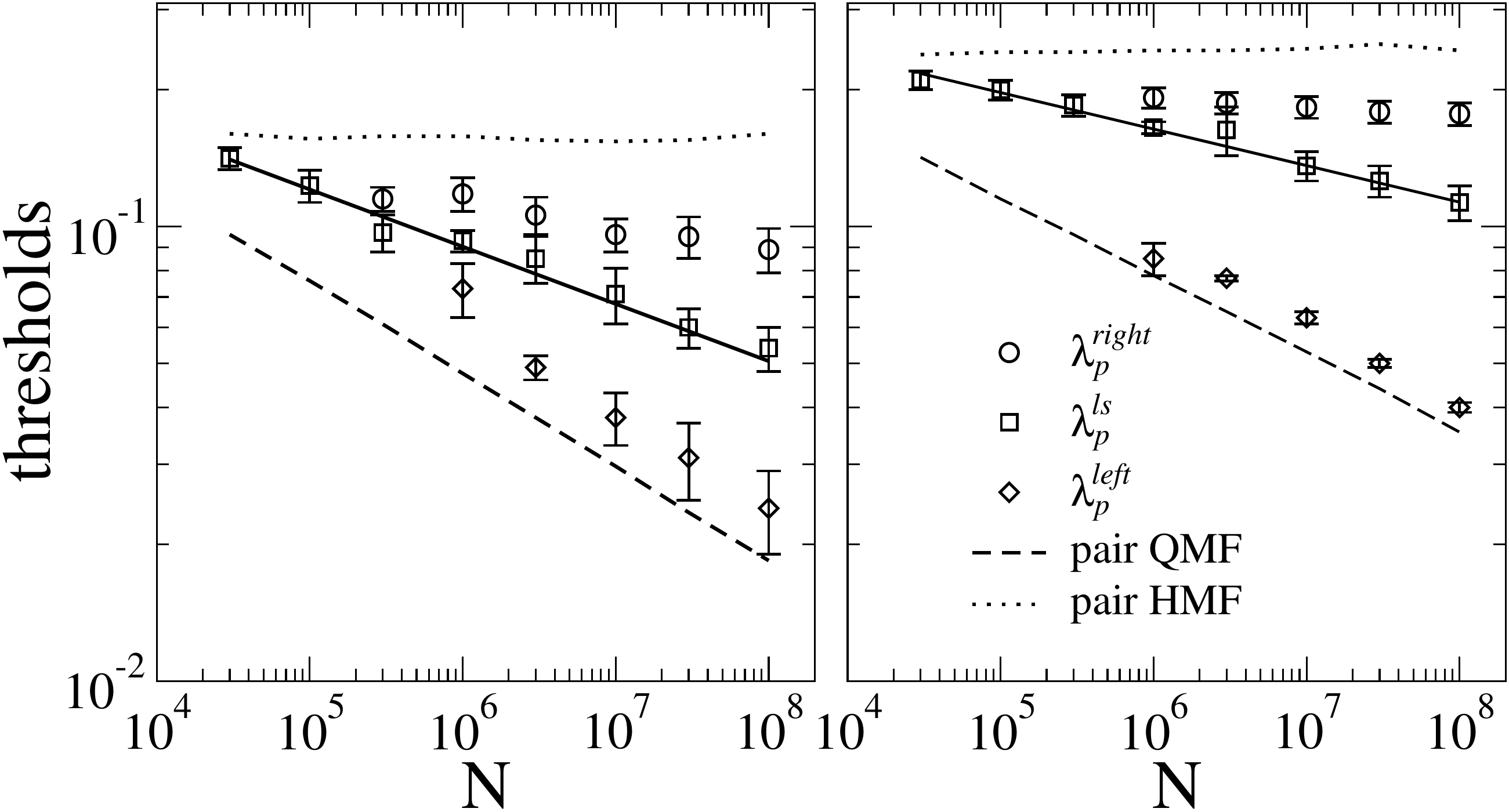}
 % lambdaN_hmf.pdf: 690x493 pixel, 72dpi, 24.34x17.39 cm, bb=0 0 690 493
 \caption{Thresholds for SIS dynamics on SF networks  with degree exponents 
$\gamma=3.5$ (left) and $\gamma=4.0$ (right). The results predicted by  pair 
QMF~\cite{mata2013pair} and  pair HMF~\cite{Mata14} theories are shown as dashed 
and doted lines, respectively. Solid lines are power law regressions. 
{Averages were done over at least 5 samples for the statistics of the
rightmost peaks and at least 20 samples for lifespan 
and leftmost peaks.}}
\label{fig:lb}
\end{figure}

Figure~\ref{fig:lb} shows the positions $\lambda_p^{left}$ (the leftmost peak),  
$\lambda_p^{right}$ and $\lambda_p^{ls}$ against the network size. One can see 
that $\lambda_p^{right}$ reaches a constant value for large $N$ whereas the 
other ones neatly decays with $N$. In a nutshell, our results show that the case 
$\gamma>3$ may concomitantly exhibit  transitions predicted by three competing 
mean-field theories: ({\it i}) At $\lambda = \lambda_p^{left}$, one has a 
transition to an epidemics highly concentrated at the star subgraph containing 
the most connected  vertex and its nearest neighbors. The threshold dependence 
on size is very well described by QMF 
theories~\cite{Castellano10,Ferreira12,mata2013pair}. ({\it ii}) At 
$\lambda=\lambda_p^{ls}$, a transition with a threshold described by the BCPS 
theory~\cite{boguna2013nature} is observed but, our numerics indicate that it is 
not endemic since PR and $\rho$ decays with $N$ above this threshold. Notice 
that the  threshold $\lambda_p^{ls}$ decays with $N$ much  slower than 
$\lambda_p^{left}$. {This interval is characterized by the 
mutual activation of stars sub-graphs centered on the outliers by means of 
reinfection mechanism proposed in the BCSP theory~\cite{boguna2013nature}}.
({\it iii}) For $\lambda=\lambda_p^{right}$, a transition 
involving an authentic endemic phase with a finite threshold is observed as 
formerly, and now surprisingly,  predicted by the HMF 
theory~\cite{Pastor01,*Pastor01b}. {Here, the bulk of the network
acts collectively in the epidemic spreading through the whole network characterizing
a real phase transition.}

\begin{figure}[hbt]
 \centering
 \includegraphics[width=3.7cm,angle=90]{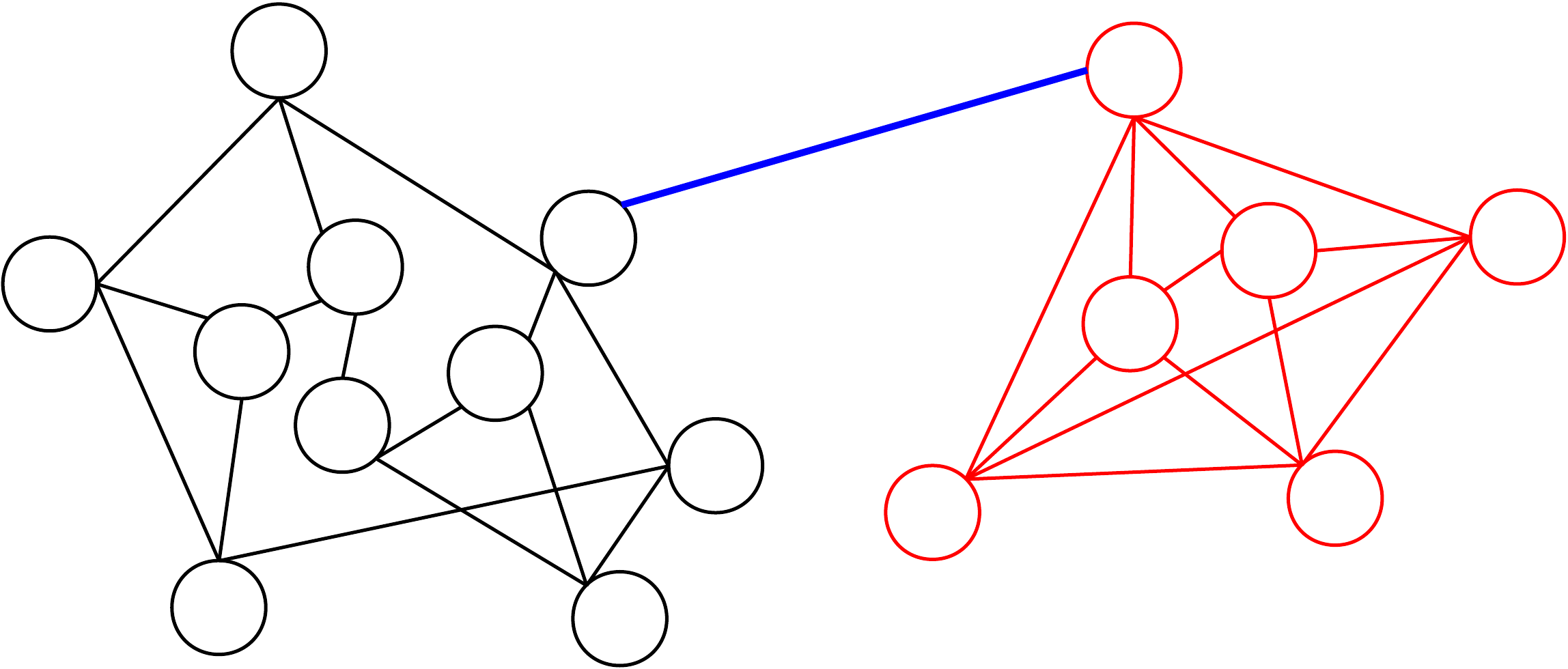}
\includegraphics[width=5.5cm]{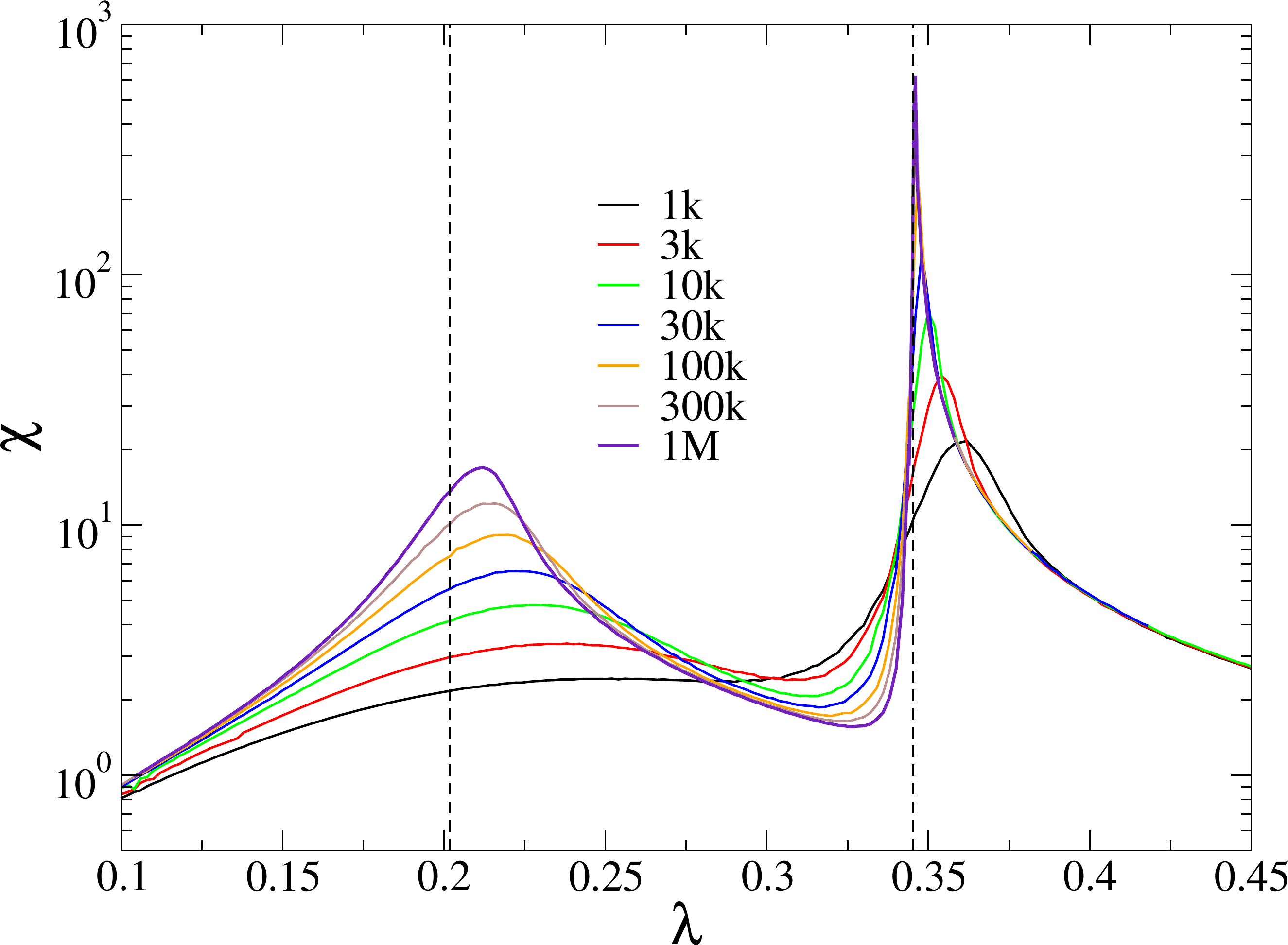}
 \caption{Left: Schematics of a double random regular network (DRRN). Right:
Susceptibility  against infection rate for DRRNs with using $m_1=4$, $m_2=6$,
$\alpha=1/2$ and different sizes. Dashed lines are thresholds predicted for
DRRNs.}
 \label{fig:rrn}
\end{figure}

The co-existence of localized and endemic transitions in a same network can be 
explained in a double random regular network (DRRN), Fig.~\ref{fig:rrn}. These 
networks are formed by two random regular networks (RRNs)\footnote{In a single 
RRN all vertices have the same degree $m$ but connections are done at random 
avoiding multiple and self  connections~\cite{Ferreira12}.} of sizes $N_1$ and 
$N_2=N_1^\alpha$ ($\alpha<1\Rightarrow N_2/N_1\rightarrow 0$ in the 
thermodynamical limit) and degree $m_1$ and $m_2$, respectively, connected by a 
single edge. The DRRN has two epidemic thresholds corresponding to the 
activations of single RRNs. Choosing $m_1=4$ and $m_2=6$, the thresholds 
determined for single RRNs are $\lambda_c^{(1)}=0.31452$ ($m_1=4$, present work) 
and $\lambda_c^{(2)}=0.2026$ ($m_2=6$~\cite{RonanEPJB}). By construction, the 
former involves an endemic and latter a localized transition since the 
smaller RRN constitutes itself a vanishing fraction of the whole network. 
Figure~\ref{fig:rrn} shows the susceptibility plots for $\alpha=0.5$ with peaks 
converging exactly to  the expected thresholds. The threshold obtained via 
lifespan method, which is in principle fitted  by the BCPS theory, converges to 
the localized one (see appendix A for additional data and discussions). This 
network model can be generalized to produce an arbitrary number of transitions 
providing a clearer analogy with multiple transitions observed for random networks
with $\gamma>3$.

{An additional property can be derived for random networks with $\gamma>3$: 
outliers have negligibly low probability to be connected to each other. Due to 
the absence of degree correlation, the probability that a vertex of degree $k$ 
is connected to an outlier of degree $k_{out}$ is given by 
$P(k|k_{out})=kP(k)/\lrangle{k}$~\cite{mariancutofss} irrespective of the outlier's 
degree. Therefore the probability that an outlier is connected to other outlier 
is given by \[P_{out} \simeq \int_{k\gtrsim\lrangle{k_{max}}} P(k|k_{out}) dk 
\sim \lrangle{k_{max}}^{-\gamma+2},\] which goes to zero for large networks 
permitting the formation of non directly connected domains centered on the 
outliers. This conclusion can be obtained rigorously using hidden variable 
formalism~\cite{Serrano}.}
We have now a simple physical explanation for multiple thresholds  
and its connection with the lifespan simulation method: 
The core containing the outliers plus its nearest neighbors form a subgraph 
with $N_2\sim \sum_{k>\lrangle{k_{max}}} NP(k)k \sim N^{1/(\gamma-1)}\ll N$. 
This domain size diverges as network increases and is able to sustain a 
long-term epidemic activity, but still represents a vanishing fraction of the 
whole  network. Above the activation of this domain but still below the endemic 
phase, the epidemics is eventually transmitted to any other vertex of the 
network due to the small-world property, but this activity dies out quickly 
outside this core since there the  process is locally sub-critical. However, all 
network vertices will be infected for some while since the active  central core 
acts as a reservoir of infectiousness to the rest of the network. 

\begin{figure}[ht]
 \includegraphics[width=7.5cm]{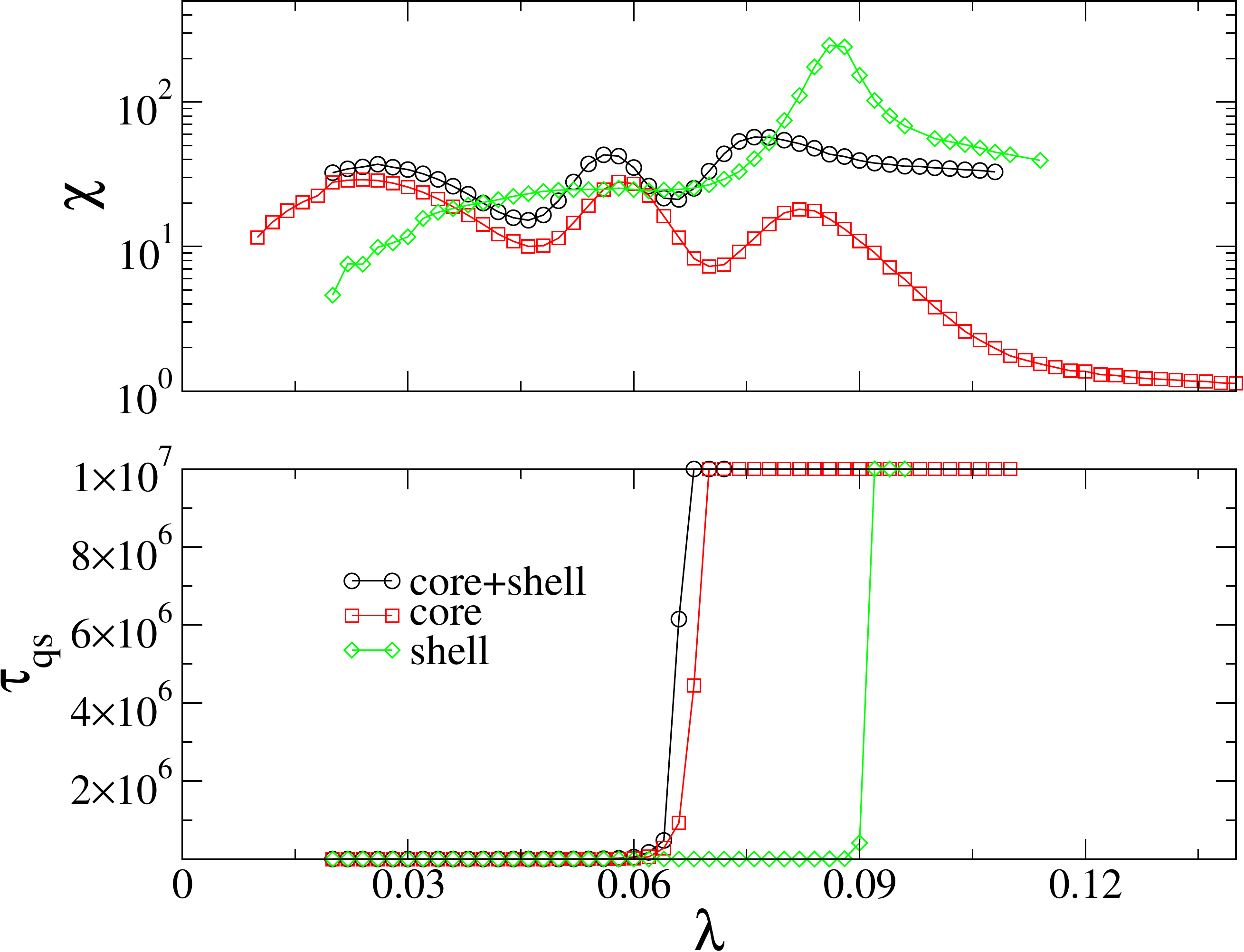}
 % suscoreg350N10.pdf: 754x577 pixel, 72dpi, 26.60x20.36 cm, bb=0 0 754 577
 \caption{(Color online) Susceptibility (top) and QS lifespan (bottom) against infection rate 
for SIS dynamics on a network with  $N=3\times 10^7$, $k_0=3$ and degree exponent 
$\gamma=3.50$ restricted to different domains (see text for definitions). The lifespan is considered 
infinite if greater than the averaging time $t_{av}=10^7$.}
 \label{fig:core}
\end{figure}

Our conjecture is confirmed in Fig.~\ref{fig:core} where SIS dynamics in a large 
network ($N=3\times 10^7$ vertices) is compared with the dynamics restricted to 
either its core of outliers (7 most connected vertices) plus their nearest 
neighbors ($\approx13200$ vertices) or to its outer shell excluding the 
core\footnote{To restrict the epidemics to the core we immunize the shell and 
vice-versa.}. The multiple peaks for the core are observed approximately at the 
same places as those for the whole network but the outer shell  exhibits a single 
peak around $\lambda_p^{right}$. However, the lifespan determined via QS method 
(see section~\ref{sec:simulations}) diverges at $\lambda\approx \lambda_p^{ls}$  
for both core and whole network whereas the divergence coincides with 
$\lambda_p^{right}$ for the outer shell.

{
We also analyzed the lifespan using the QS method, Eq.~(\ref{eq:tauqs}), for a 
fixed distance to both leftmost and lifespan peaks. For the investigated size 
range $N<10^8$, the lifespan values are relatively short ($<10^2$) and increase 
algebraically with system size   in the interval 
$\lambda_p^{left}<\lambda<\lambda_p^{ls}$ while long and exponentially diverging 
lifespans, granting long-term activity for large networks~\cite{Ganesh05}, are 
obtained for $\lambda_p^{ls}<\lambda<\lambda_p^{right}$. The algebraic 
dependence for the former case is almost certainly a finite-size effect. We 
calculated the lifespan for $\lambda_p^{left}<\lambda<\lambda_p^{ls}$ for the SIS model on star 
graphs with $k$ leaves and an algebraic growth of the lifespan with $N$ is also obtained 
for $k<2000$ which coincides with the range size of typical star subgraphs 
obtained for UCM networks investigated here. However, a crossover to an 
exponential growth is obtained for larger star graphs ($k>10^4$) showing that 
this structure is itself able to sustain alone a long-term epidemic activity. So, if 
one could simulate SIS model on much larger UCM networks ($N > 10^{12}$) the 
threshold $\lambda_p^{left}$ would define a transition to a localized but 
long-term epidemics and the lifespan method would detect the transition given by
the QMF theory.
}

\begin{figure}[ht]
 \centering
\includegraphics[width=8.5cm]{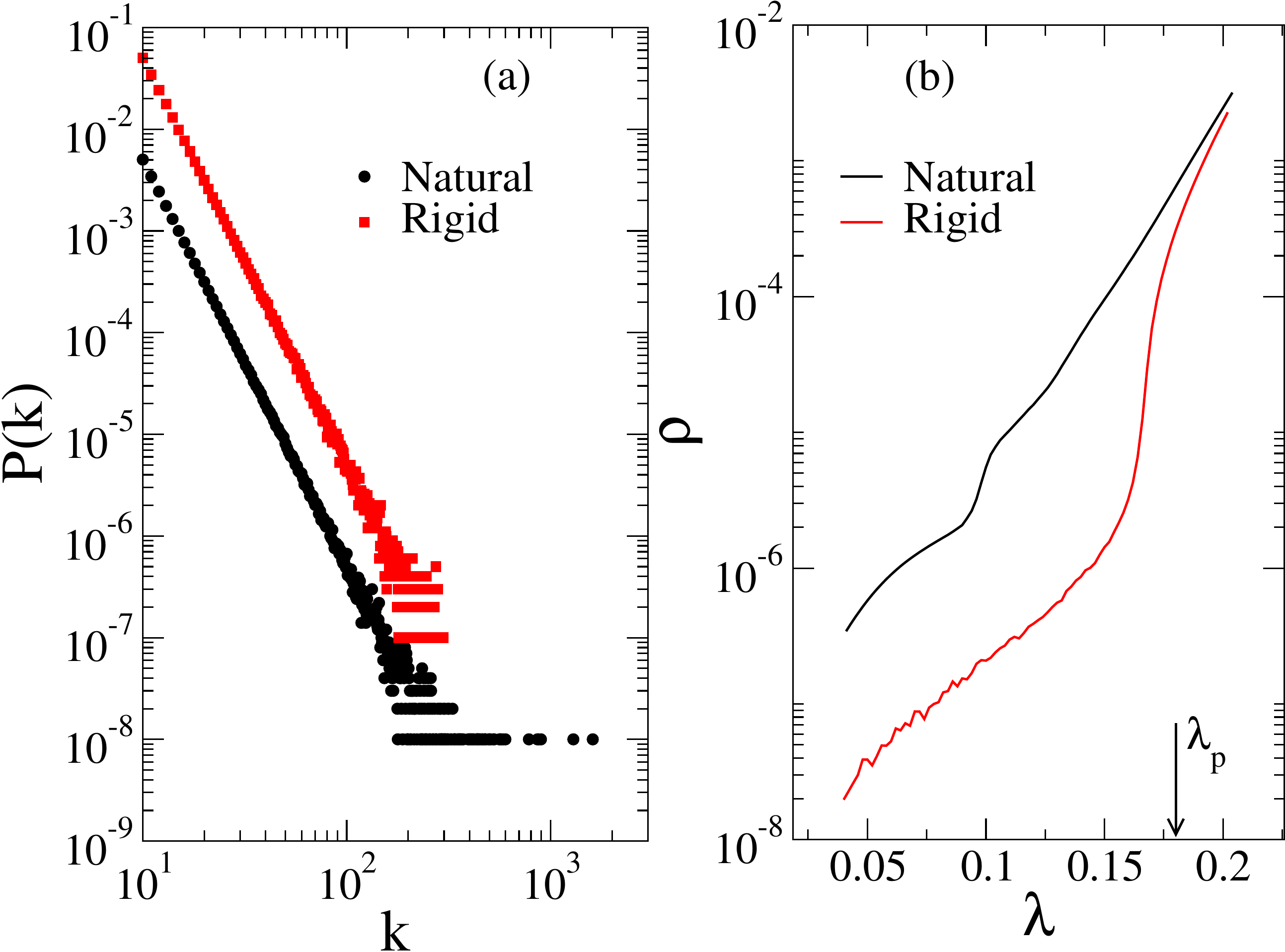}
%\subfigure[][]{\includegraphics[width=5.5cm]{./rho_rig_natN11.pdf}}
 \caption{(Color online) \label{fig:hard}  (a) The tail of the degree distributions
for networks with $\gamma=4$, $k_0=3$, and $N=10^8$ vertices and either rigid or natural cutoff.
The curve for rigid cutoff was shifted to enhance visibility. (b) QS density against
infection rate for a network degree exponent $\gamma=4.0$ using different
cutoffs.}
\end{figure}

\begin{figure}[ht]
 \centering
%\subfigure[][]{\includegraphics[width=5.5cm]{./sus_cutoff_g400eg350.pdf}}
{\includegraphics[width=8.5cm]{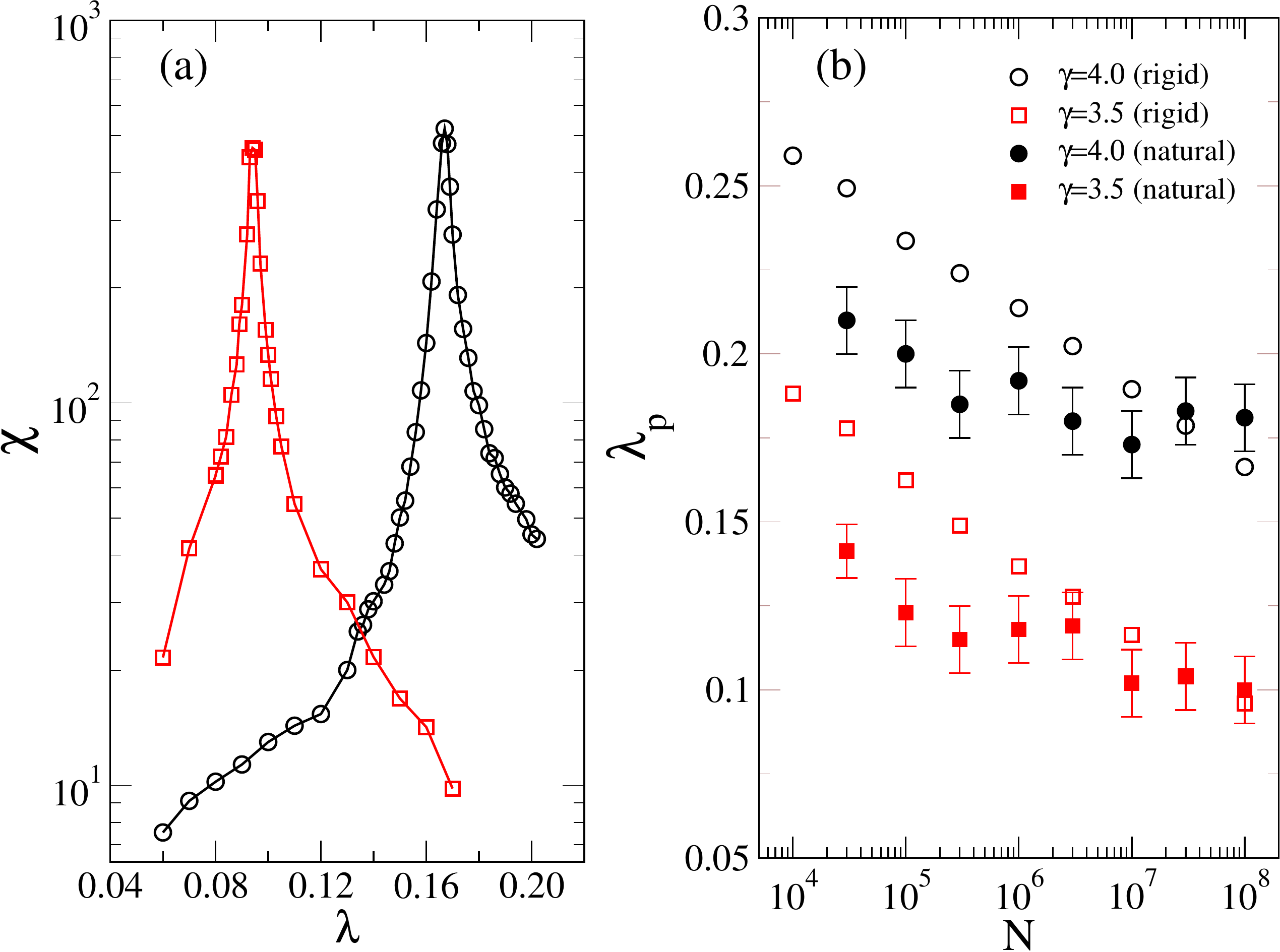}}
 \caption{\label{fig:hard2}  (Color online) (a) Susceptibility curves for two networks with rigid 
cutoff, $k_0=3$,  $N=10^8$ vertices and different degree exponent (symbols are the same 
used in panel (b)). (b) Threshold against system size for rigid and natural 
cutoffs. {The averages were done over at least 6 samples for rigid cutoff but error bars
are smaller than symbols. Averages for natural cutoffs are the same as in Fig.~\ref{fig:lb}.}}
\end{figure}

Outliers play a central role even not being able to produce separately a genuine 
endemic phase where the whole network has a non-vanishing probability of being 
infected. To highlight such a role,  we introduce a hard cutoff  in the degree 
distribution as $k_{max}=k_0N^{0.75/(\gamma-1)}$, which suppresses the emergence of 
outliers as shown in Fig.~\ref{fig:hard}(a). This choice is because random 
networks without a rigid upper bound have a highly fluctuating natural cutoff,
as discussed above. Fig.~\ref{fig:hard}(b) compares the 
QS density for rigid and natural cutoffs. The infectiousness for 
$\lambda<\lambda_c^{endemic}$ is highly reduced in the absence of outliers. The 
susceptibility no longer exhibits multiple peaks for a hard cutoff, as can be 
seen in Fig.~\ref{fig:hard2}(a), confirming the existence of a single 
transition. Also, the  thresholds for hard cutoff networks are quite close  to 
$\lambda_p^{right}$ obtained with the natural cutoff, as shown in 
Fig.~\ref{fig:hard2}(b). Such an observation is in agreement with the HMF theory 
where the thresholds for $\gamma>3$ are  asymptotically independent of how $k_c$ 
diverges~\cite{Pastor01,*Pastor01b,Mata14}. 

\vspace{-0.5cm}
\section{Conclusions}
\label{sec:conclusions}
In summary, we thoroughly simulated the dynamics of the SIS epidemic model on 
complex networks with power law degree distributions with exponent $\gamma>3$, 
for which conflicting theories discussing the existence or not of a finite 
epidemic threshold for the endemic phase have recently been 
proposed~\cite{Castellano10,Goltsev12,boguna2013nature,Lee2013}. We show that 
the SIS dynamics can indeed exhibit several transitions associated to different 
epidemiological scenarios. Our simulations support a  picture where the 
threshold obtained recently in the BCPS mean-field 
theory~\cite{boguna2013nature} represents a transition to  localized epidemics 
in random networks wih $\gamma>3$ and that the transition to an authentic 
endemic state, in which a finite fraction of network is infected, possibly 
occurs at a finite threshold as formerly and now surprisingly foreseen by the 
HMF  theory~\cite{Pastor01,*Pastor01b}. The multiple transitions are associated 
to large gaps in the degree distribution \revis{with a few outliers}, which 
permits the formation of non-directly connected domains of activity centered on 
these outliers. \revis{If the number of hubs is large, as in the case of SF 
networks with $\gamma<3$, every vertex of the network is ``near'' to some hub 
and the activation of hubs implies in the activation of the whole network, as 
previously reported in \cite{Ferreira12,mata2013pair}.} Our finds  are 
consistent with the conjecture proposed by Lee \textit{et al}.~\cite{Lee2013} 
since the lifespans of independent domains involving outliers grow exponentially 
fast with the domain sizes implying that long-term epidemic activity is possible 
even in non-endemic phase. Our finds also do not rule out the mean-field 
analysis of Ref.~\cite{Goltsev12}. The intermediary transitions can be 
associated to distinct localized eigenvectors that are centered on the outliers 
while the endemic threshold involves a delocalized eigenvector with a finite 
eigenvalue. 

{Our results are in consonance with a recent line of investigation, in which the 
topological disorder in networks with heterogeneous degree  distribution may 
produce rare regions and  Griffiths phases leading to anomalous behaviors in the 
subcritical phase~\cite{Lee2013,GP,Buono,odor2013spectral}. Such  anomaly is 
characterized by localized  activity that survives for long  times, even though 
the network is macroscopically absorbing. Very recently, the possibility of rare 
regions effects from pure topological disorder in the SIS dynamics on unweighted 
SF networks as well as  multiple  transitions were suggested in 
Ref.~\cite{odor2014localization}. Our results may, thus, be a fingerprint of 
GPs. However, more detailed analyses are demanded for a conclusive relation.} 
Also, very recently, multiple phase transitions were found in  percolation 
problems on SF networks with high clustering~\cite{colomer2014} and on networks 
of networks~\cite{bianconi2014multiple}. In both cases transitions were  
hallmarked by multiple singular points  in the order parameter in analogy with 
our results for epidemics.

Our final overview is that apparently competing mean-field 
theories~\cite{Pastor01,*Pastor01b,Castellano10,Goltsev12,boguna2013nature, 
Lee2013} can be considered, in fact,  complementary, describing distinct 
transitions that may concomitantly emerge depending on the network structure. In 
particular, the transitions involving localized phases, as possibly the one 
predicted by the BCPS theory~\cite{boguna2013nature}, are not  
negligible since they become long-term and an epidemic outbreak may eventually 
visit a finite fraction of the network. This peculiar result is unthinkable for 
other substrates rather than complex networks sharing the small-world and 
scale-free properties. Actually, it is well known that some computer viruses can 
survive for long periods (years) in a very low density (below 
$10^{-4}$)~\cite{romuvespibook}, exemplifying the importance of metastable 
non-endemic states. Our numerical results call for general theoretical 
approaches to describe in an unified framework the multiple transitions of the 
SIS dynamics on SF networks.

\vspace{-0.5cm}
\begin{acknowledgments}
This work was partially supported by the Brazilian agencies CAPES, CNPq and 
FAPEMIG. Authors thank Romualdo Pastor-Satorras, Claudio Castellano and Mari\'an 
Bogu\~{n}\'a for the critical and profitable discussions and Ronald Dickman for 
priceless suggestions.
\end{acknowledgments}
\vspace{-0.5cm}
\appendix
\section{Quasistationary versus lifespan methods}
\label{sec:appendix}

\begin{figure}[hbt]
 \centering
\includegraphics[width=8.5cm]{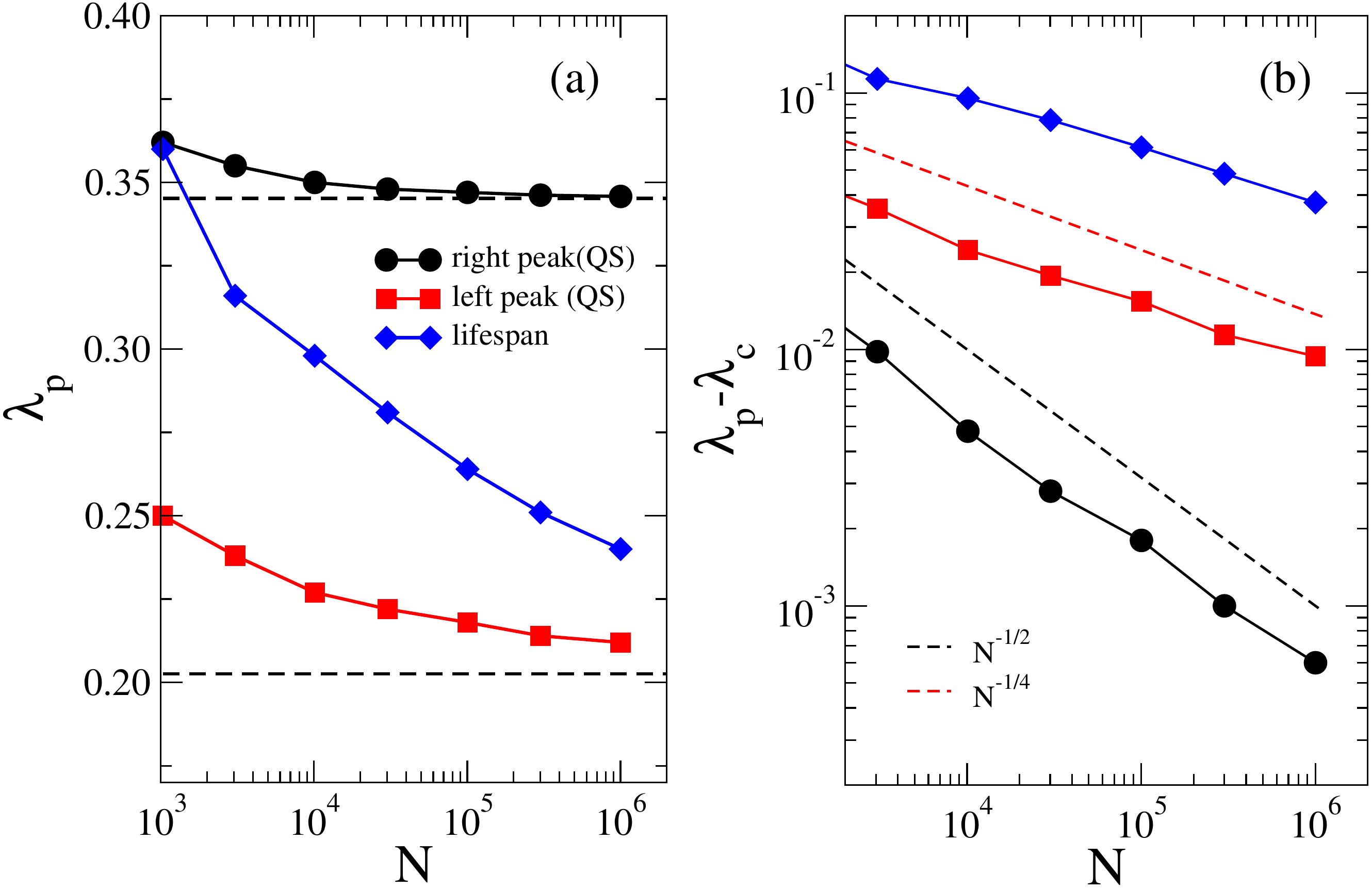}
%\subfigure[][]{\includegraphics[width=5.7cm]{./thres_2rrn_dif.pdf}}
\caption{(Color online) Threshold analysis for DRRN with $\alpha=1/2$, $m_1=4$, and 
$m_2=6$. (a) The thresholds estimated as the peaks in the susceptibility or 
lifespan curves. The dashed lines are thresholds obtained on single RRNs with 
the respective $m_i$. (b) Difference between peaks and the thresholds for single 
RRNs with $m=4$ (lifespan and left susceptibility peaks) or $m=6$ (right 
susceptibility peak).}
 \label{fig:2rrn_sim2}
\end{figure}

\begin{figure}[hbt]
 \centering
\includegraphics[width=8.5cm]{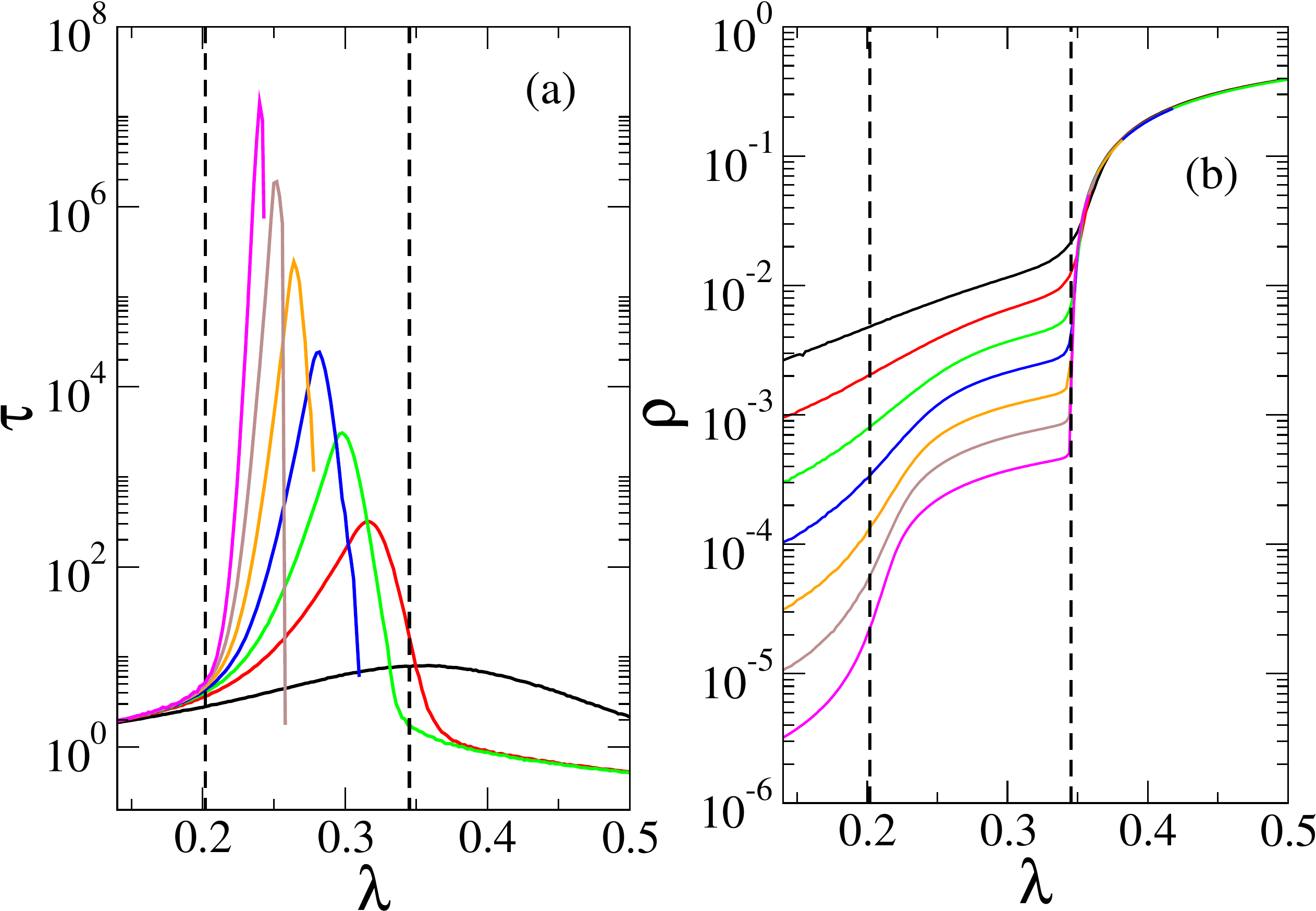}
%\subfigure[][]{\includegraphics[width=4.0cm]{./rho_2rrn.pdf}} \\
\caption{(Color online) (a) The lifespan according Ref.~\cite{boguna2013nature} and (b) the QS 
density of infected vertices against infection rate for SIS  model on DRRNs with 
$\alpha=1/2$, $m_1=4$,  $m_2=6$, $N_1=10^3,~3\times 10^3,~10^4,~3\times 
10^4,~10^5,~3\times10^5,$ and $10^6$  (increasing to left or bottom in (a) or (b), respectively). 
Dashed vertical lines indicate the activation thresholds in each sub-graph.}
 \label{fig:2rrn_sim}
\end{figure}

Lets show that the QS method succeeds whereas lifespan method fails in 
predicting the endemic phase for a DRRN (Fig.~\ref{fig:rrn}).  The 
susceptibility peaks in Fig.~\ref{fig:rrn} clearly converge to the respective 
thresholds of single RRNs as highlighted in Fig.~\ref{fig:2rrn_sim2}(a) and (b). 
Notice that the mean-field theory for the finite-size scaling of the contact 
process, which in the case of strictly homogeneous networks is exactly the same 
as SIS model with a rescaled infection rate $\lambda/m$, predicts  that the 
threshold approaches its asymptotic values as $\lambda_p-\lambda_c\sim 
S^{-1/2}$, where $S$ is the graph size~\cite{Mata14}. So, the endemic threshold 
is expected to scale as $\lambda_p-\lambda_c^{(1)}\sim N_1^{-1/2}\sim N^{-1/2}$ 
and the localized one as $\lambda_p-\lambda_c^{(2)}\sim N_2^{-1/2}\sim 
N^{-\alpha/2}$. These power-laws are confirmed in Fig.~\ref{fig:2rrn_sim2}(b). The 
lifespan curves, obtained using as initial condition only the most connected 
vertex infected (the one connecting sub-graphs), have single peaks that converge 
to the threshold corresponding to a localized epidemic and interestingly 
following the same scaling law as the left QS peak as shown in 
Figs.~\ref{fig:2rrn_sim2} and \ref{fig:2rrn_sim}(a). The central point here is 
that the lifespan method detected the first threshold where the absorbing state 
becomes globally unstable (an exponentially long-term activity) that,  in this 
case, is not the endemic one as shown in Fig.~\ref{fig:2rrn_sim}(b), in which the QS 
density is shown as a function of the infection rate.

It is worth noticing that the QS simulations around the peaks are orders of 
magnitude computationally more efficient than the lifespan method.

%\bibliographystyle{apsrev4-1}
%\bibliography{sis_comment}


\end{document}